\documentclass[%
 reprint,
 amsmath,amssymb,
 aps,
]{revtex4-2}

\usepackage{cprotect}
\usepackage{xcolor} 
\usepackage{graphicx}
\usepackage{dcolumn}
\usepackage{bm}
\usepackage{float}
\usepackage{setspace}
\usepackage{comment}
\usepackage{adjustbox}

\usepackage{upgreek}
\usepackage{makecell}
\usepackage{caption}
\usepackage{subcaption}

\usepackage{graphicx, float}
\usepackage{caption,lipsum}

\usepackage{stfloats} 

\begin{document}

\preprint{APS/123-QED}

\title{An Alternative Design for Large Scale Liquid Scintillator Detectors}

\author{Iwan Morton-Blake}
\author{Steven D. Biller}
\affiliation{Department of physics, University of Oxford, Oxford OX1 3RH, United Kingdom}

\date{\today}

\begin{abstract}
The construction of large-scale liquid scintillator detectors is complicated by the need to separate the scintillation region from photomultiplier tubes (PMTs) due to their intrinsic radioactivity. This is generally done using acrylic or nylon barriers, whose own activity can also lead to substantial reduction of the fiducial detection volume for a number of low energy ($\sim$MeV) studies, and whose construction becomes increasingly difficult and expensive for larger detector volumes. Here we present an initial simulation study of an alternative geometric construction known as SLiPS (Stratified Liquid Plane Scintillator), which aims to avoid such physical barriers entirely by instead using layers of lipophobic liquids to separate PMTs from the scintillation region.
\end{abstract}


\maketitle

\section{Introduction}

Large scale liquid scintillation (LS) detectors have played a central role in non-accelerator particle physics, including experiments such as KamLAND \cite{PhysRevC.85.045504}, Borexino \cite{Borexino}, Daya Bay \cite{DayaBay}, RENO \cite{RENO}, Double Chooz \cite{DoubleChooz:2019qbj}, and with several more in early operation, construction or planning, such as SNO+ \cite{SNO+}, JUNO \cite{JUNO} and THEIA \cite{THEIA}. Common to all large-scale liquid scintillator detectors is the requirement for a boundary separating the scintillation region from the photomultiplier tubes (PMTs), due to their intrinsic radioactivity. This is generally accomplished using acrylic or nylon barriers, whose own radioactivity can also lead to substantial cuts to the fiducial detection volume for a number of low energy ($\sim$MeV) studies. Such barrier constructions become increasingly difficult and expensive with increasing detector volumes. This paper uses simulation to explore an alternative geometric design for larger LS detectors aimed at simplifying such constructions while also making more economic use of PMTs to maintain high photocathode coverage. The approach opts for a planar design using lipophobic liquids to separate detection and scintillation regions without additional physical barriers, and reflective surfaces to increase light collection and reconstruct event using the time-separated wave fronts. The result is a potentially more scalable approach that may permit clean construction of larger scale instruments in the future.

\section{SLiPS Detector Design}

\begin{figure}[h]
      \centering
      \begin{subfigure}[b]{0.47\textwidth}
         \includegraphics[width=\textwidth]{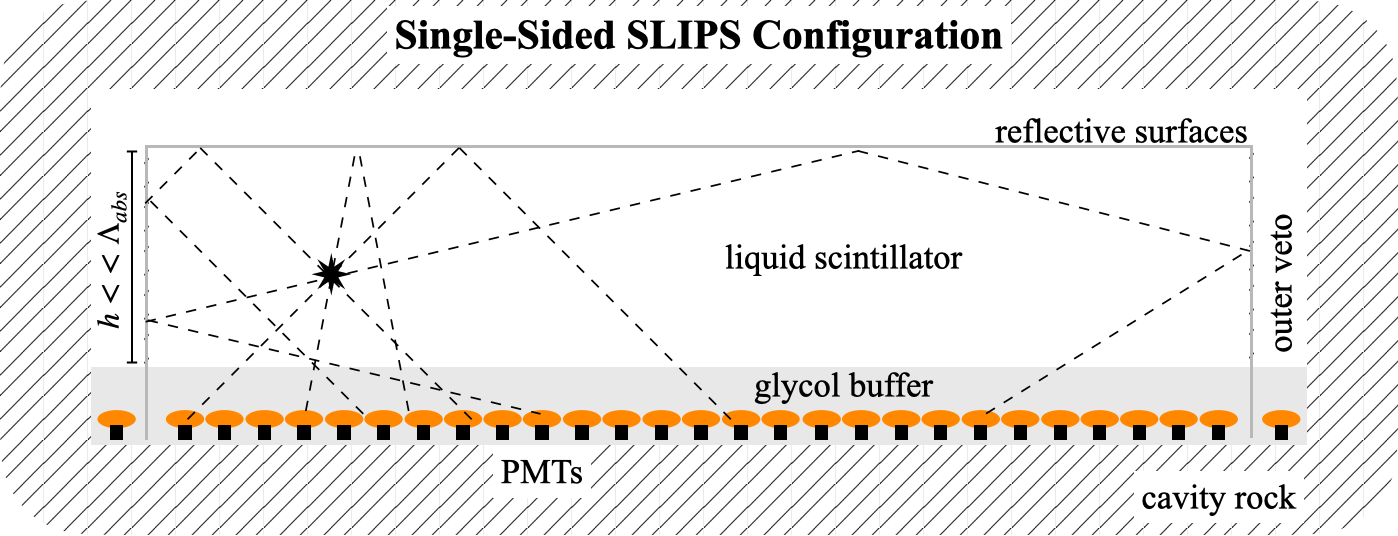}
         \caption{}
      \end{subfigure}
      \begin{subfigure}[b]{0.45\textwidth}
         \includegraphics[width=\textwidth]{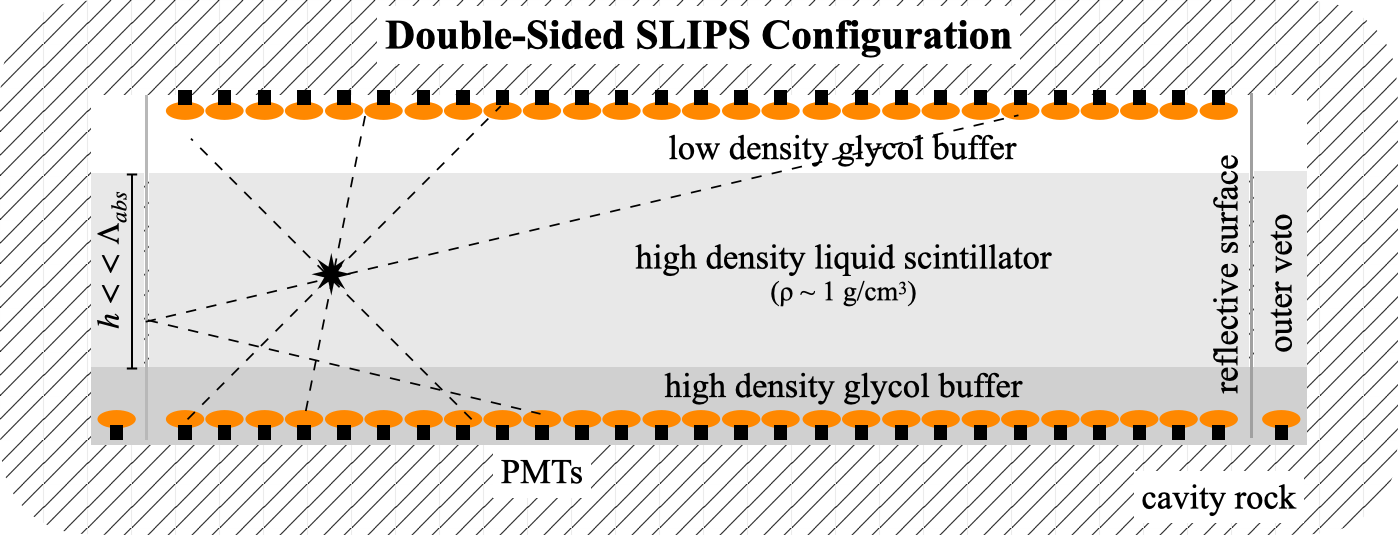}
         \caption{}
         \label{SLIPS_double}
      \end{subfigure}
     \caption{Rough schematic cross sectional views of (a) single-sided and (b) double-sided SLiPS configurations. Scintillation light paths from a hypothetical interaction are represented by dashed lines.}
	 \label{SLIPS}
\end{figure}

\begin{figure}[h]
\centering
\includegraphics[width=75mm]{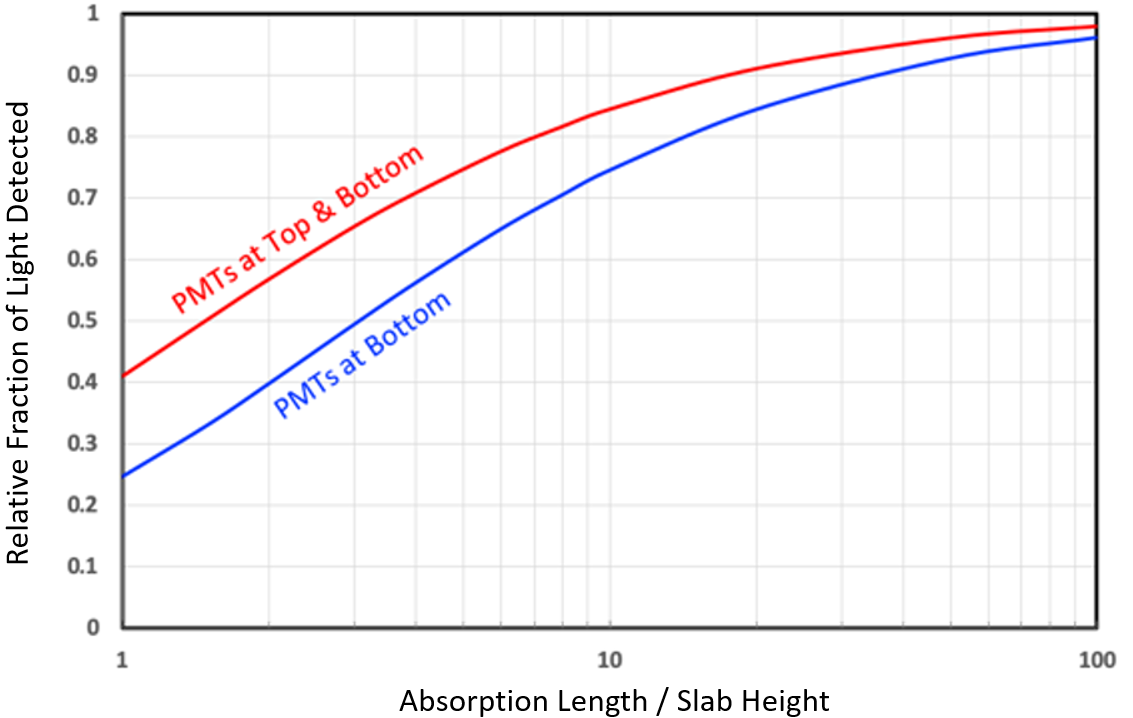}
\caption{Simplified analytical model of the fraction of produced light detected vs the ratio of absorption length to scintillator depth. Thus, a typical $\sim$40m absorption length and the 10m scintillator depth considered here would correspond to a ratio of 4 on this plot.}
\label{analytic_light}
\end{figure}

Figure \ref{SLIPS} shows a sketch of the Stratified Liquid Plane Scintillator (SLiPS) detector for two configurations. The surrounding rock cavity depicted is only notional, as it would more realistically be engineered with a curved roof etc. to provide structural integrity, but this is not relevant for the current study. PMTs are mounted on the bottom of a wide cavity and submerged in a distillable, lipophobic liquid, above which a less dense scintillator is layered \cite{Morton-Blake}. In many ways, this layering is similar to the ``partial fill'' configuration of the SNO+ detector, which ran very successfully for nearly a year with scintillator floating above a water interface \cite{Morton-Blake}. During this time, there was no evidence of instabilities or mixing near the interface, even following re-circulation or filling activities, no evidence of fluor (PPO) transfer from scintillator to water, and no evidence of U/Th transfer from water to scintillator. Event reconstruction was accomplished using fitting algorithms specially adapted to account for the two media, and the calibration and analysis of data in both scintillator and water regions was successfully performed. Small scale tests with LAB and ethylene glycol also confirm no significant transfer of fluors between these liquids, even when agitated and allowed to settle.

Ideally, the refractive indices of the two liquids should be matched as closely as possible in the wavelength range of emission, which is typically $\sim$400-450nm. For this reason, buffer liquids such as ethylene glycol are good candidates. For the single-sided configuration, reflective surfaces near the top and sides of the detector are used to direct scintillation light down to the bottom PMT array. These reflective sheets might also be used to provide optical shielding from light emitted by radiation outside the fiducial volume. PMTs may also be placed in the region between the reflective surfaces and the container to serve as a muon veto. In principle, it is also possible to construct a detector with PMT planes on both top and bottom, as in figure \ref{SLIPS}b, by using a solvent with a high enough density to allow for a secondary buffer such that $\rho$(buffer~1)$>\rho$(solvent)$> \rho$(buffer~2). However, the studies here will largely concentrate on exploring the use of reflections in the two-liquid geometry, schematically depicted in figure \ref{SLIPS}a.

\begin{figure}[h]
	\centering
	\includegraphics[width=68mm]{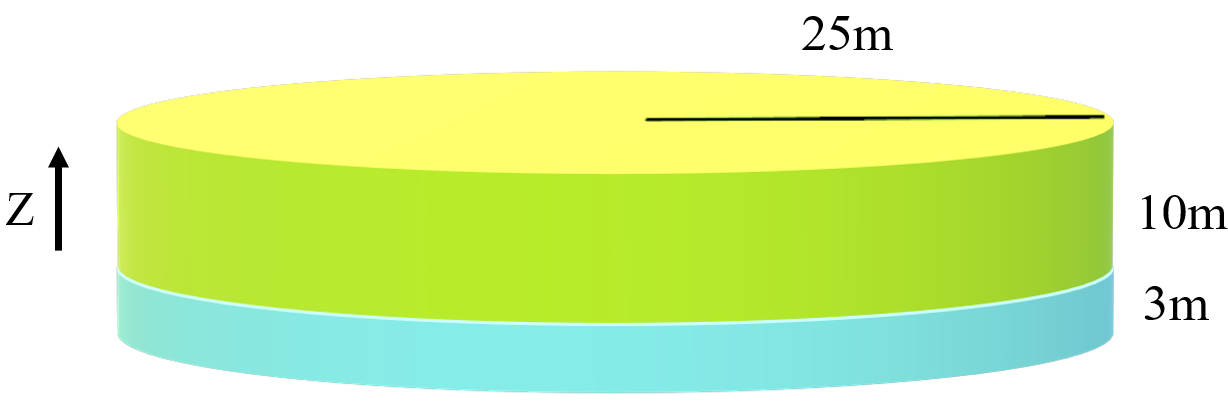}
	\caption{SLiPS scintillator and buffer dimensions used in simulations tests, and the defined z-axis direction.}
	\label{PancakeDimensions}
\end{figure}

Due to the long path lengths of scintillation light created by reflections, the detected light levels can become limited by the extinction length of photons in the scintillator. Figure \ref{analytic_light} shows results of a simple analytical calculation for the relative light levels observed in an idealised geometry involving either one or two infinite detection planes as a function of the ratio of the scintillator absorption length to scintillator plane thickness. It is therefore advantageous for the detector height to be much less (by a factor of $\sim$10) than the extinction length, generally leading to a short and wide detector configuration. This also tends to simplify the nature of the reflections for the single plane case. While alternative detector shapes are explored later, the baseline configuration for this study will be an azimuthally symmetric circular `Pancake' detector with reflective top and side surfaces shown in figure~\ref{PancakeDimensions}.

\section{Detector Simulation}
The propagation of radiation in the detector was carried out using \texttt{GEANT4} \cite{Geant4ToolKit} based simulation. The \texttt{GLG4Sim} package was used to generate energy deposition in the scintillator and the subsequent light emission \cite{GLG4Sim}. PMT and data acquisition response was also simulated assuming that the full wave-train was recorded to take advantage of information from multiple hits on individual PMTs. In practice, this only has modest advantage over using the earliest hit times for the energy range explored here, as PMT occupancies tend to be below 10\%, but may be more useful for higher energy events.

\subsection{PMTs}

Two PMT types were simulated in the performance testing presented in this work: the 20" R12860 PMT from Hamamatsu Photonics and the 8" R5912-100 Hamamatsu model. The 20" PMTs were modeled in simulation based on the specifications according to \cite{20InchPMTs}, and the 8" PMTs based on \cite{HamamatsuCatalogue} and \cite{kaptanoglu_5912mod_2018}, where the timing measurements of an un-modified R5912-100 were taken from the latter (which yielded a faster TTS than the standard data sheet). \texttt{GEANT4} geometry models of the PMT's dimensions were created and the measured quantum efficiency spectra was attributed to the photocathode surface. A collection efficiency of 90\% was assumed. The measured single photoelecton (SPE) charge spectrum and the transit time spread (TTS) were also included in the simulation. The charge spectrum was modeled as a Gaussian distribution, from which PMT hit charges were sampled for the triggering of PMTs and the global trigger. Table \ref{Table_PMTs} summarises the key parameters used to characterise each PMT.

\begin{table}[H]
\caption{Key parameters assumed in the simulation of the two types of PMTs presented in this work.} 
\begin{center}
	\begin{tabular}{||c c c||}
		\hline
		Parameter & r12860 (20") & r5912-100 (8")\\
		\hline\hline
		Peak to Valley Ratio & 4.75 & 4.2 \\ 
		\hline
		\makecell{TTS ($\upsigma$)} & 1.3ns & 0.87ns \\
		\hline
		\makecell{QE (400nm)}& 31$\%$ & 32$\%$ \\ 
		\hline
\end{tabular}
\end{center}
\label{Table_PMTs}
\end{table}

Channel discriminator thresholds were set such that, for the r12860 PMT, a photon that produced a single p.e. on the photocathode would have a trigger inefficiency of 3$\%$ (4$\%$ for the r5912-100 PMT). A trigger gate event width of 400ns was assumed. The PMTs were distributed in a hexagonal packing planar arrangement, with 54cm or 20.8cm center-to-center between adjacent tubes for 20" and 8" tubes, respectively. In all simulations, the PMTs were positioned such that the top of the photocathode cap was 2m below the scintillator-buffer interface so as to shield the scintillator from the PMT's intrinsic radioactivity.

\subsection{Simulated Liquid Scintillator and Buffer Liquid}

\begin{figure}[h]
      \centering
      \begin{subfigure}[b]{0.47\textwidth}
         \includegraphics[width=\textwidth]{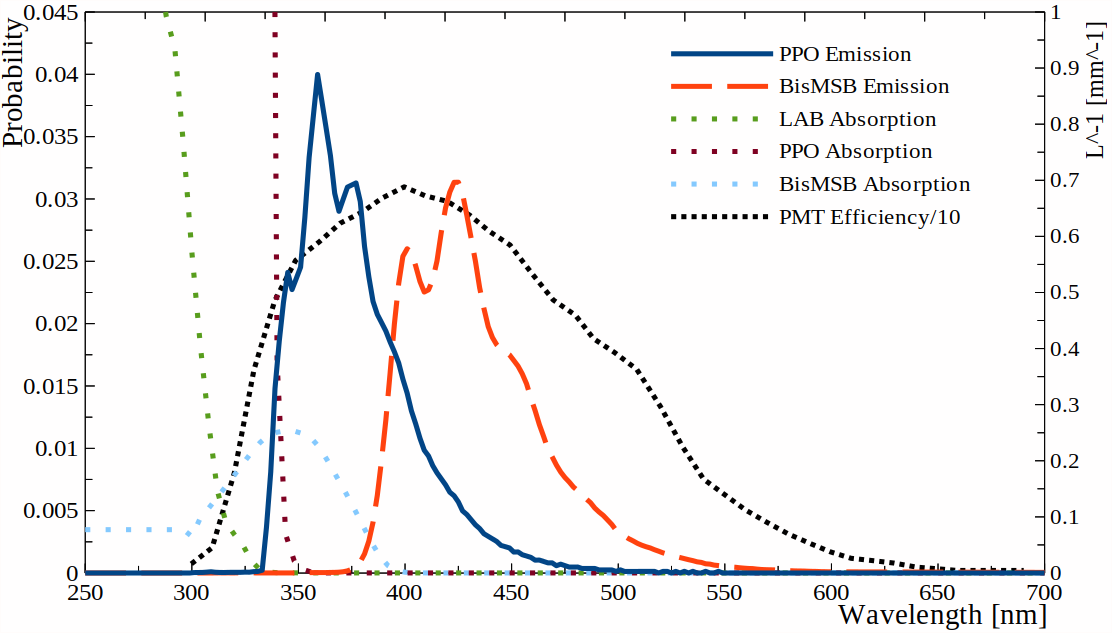}
         \caption{}
      \end{subfigure}
      \begin{subfigure}[b]{0.45\textwidth}
         \includegraphics[width=\textwidth]{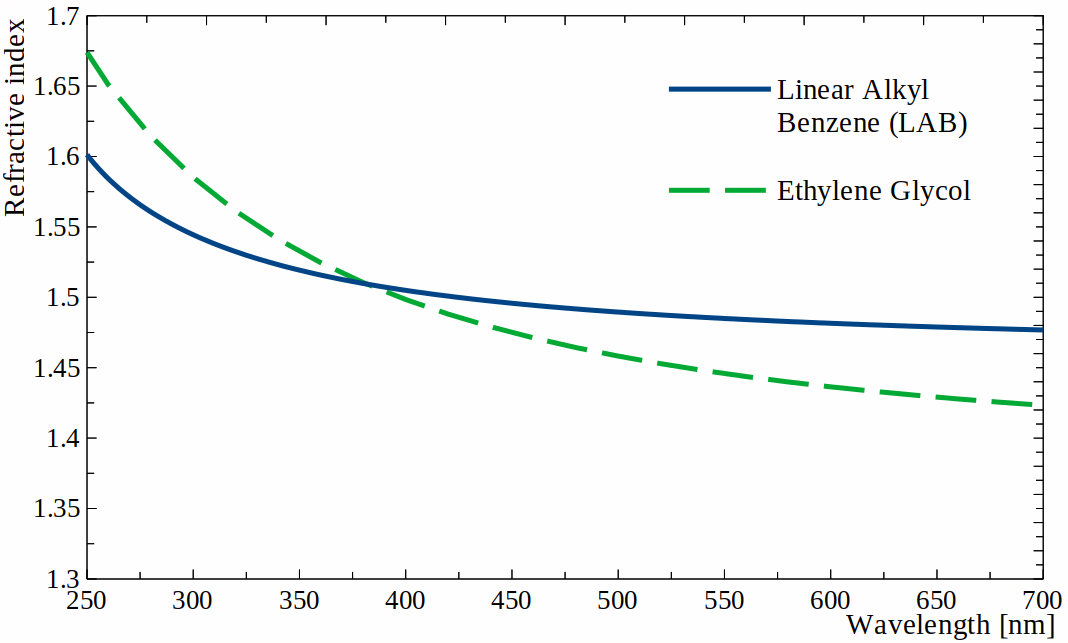}
         \caption{}
         \label{LAB_EGlycol_Rindex}
      \end{subfigure}
     \caption{(a) Absorption and Emission spectra for the LAB, PPO and bisMSB components in the scintillator cocktail. The left y-axis shows probability of emission. The black dashed line shows the r12860 Hamamatsu 20" PMT measured total detection efficiency divided by 10. The right y-axis corresponds to the sparse dotted graphs showing absorption in units of mm\textsuperscript{-1} (b) Refractive index as a function of wavelength  for LAB and ethylene glycol \cite{SNO+Scintillator}\cite{EG_rindex_paper}.}
	 \label{PMTQEScintCocktailEmissionAbsorption}
\end{figure}

The detector's design principle requires the use of immiscible liquids with similar refractive indices. Since most organic scintillator solvents tend to be hydrophobic, these can therefore be paired with hydrophilic liquids, such as glycols, which tend to be more dense. This then more naturally leads to a design with scintillator on top and PMTs in a bottom buffer region. The scintillator mixture adopted for simulation was composed of linear alkylbenzene (LAB) as the solvent plus 2g/L of 2,5-diphenyloxazole (PPO) as the primary fluor and 15mg/L of 1,4-bis-methylstyryl-benzene (bis-MSB) as the secondary fluor \cite{SNO+Scintillator}. Figure \ref{PMTQEScintCocktailEmissionAbsorption} shows the assumed absorption and emission spectra of the contributing components present in the scintillator.

Ethylene glycol was chosen as the buffer liquid to pair with LAB for this study. Ethylene glycol can be purified through distillation as well as filtration and has a similar refractive index to LAB in the wavelength region of interest, as shown in figure \ref{LAB_EGlycol_Rindex}. Matching refractive indices in the buffer and scintillator liquids minimises the refraction effects for light crossing their interface, improving light detection efficiency and simplifying event reconstruction. The scintillator solution and ethylene glycol were assumed to be 20$^\circ$C, with densities 0.858 and 1.114g/cm\textsuperscript{3}, respectively.

\section{Performance Tests}

The 'Pancake' design explored here consists of a cylindrical scintillator volume of thickness 10m and 25m in radius. The glycol buffer region extends 3m below the scintillation layer, as shown in figure \ref{PancakeDimensions}. The walls and ceiling surrounding the scintillator region are assumed to have a specular reflection efficiency of 90$\%$. The center of the cylindrical scintillator volume is defined as the origin and the x-y plane was aligned with the plane of the PMTs, which resides at the bottom of the glycol buffer with the photocathode surfaces facing upwards, 2m below the scintillator interface.

\subsection{Light Collection}
\begin{figure}[H]
	\centering
	\includegraphics[width=75mm]{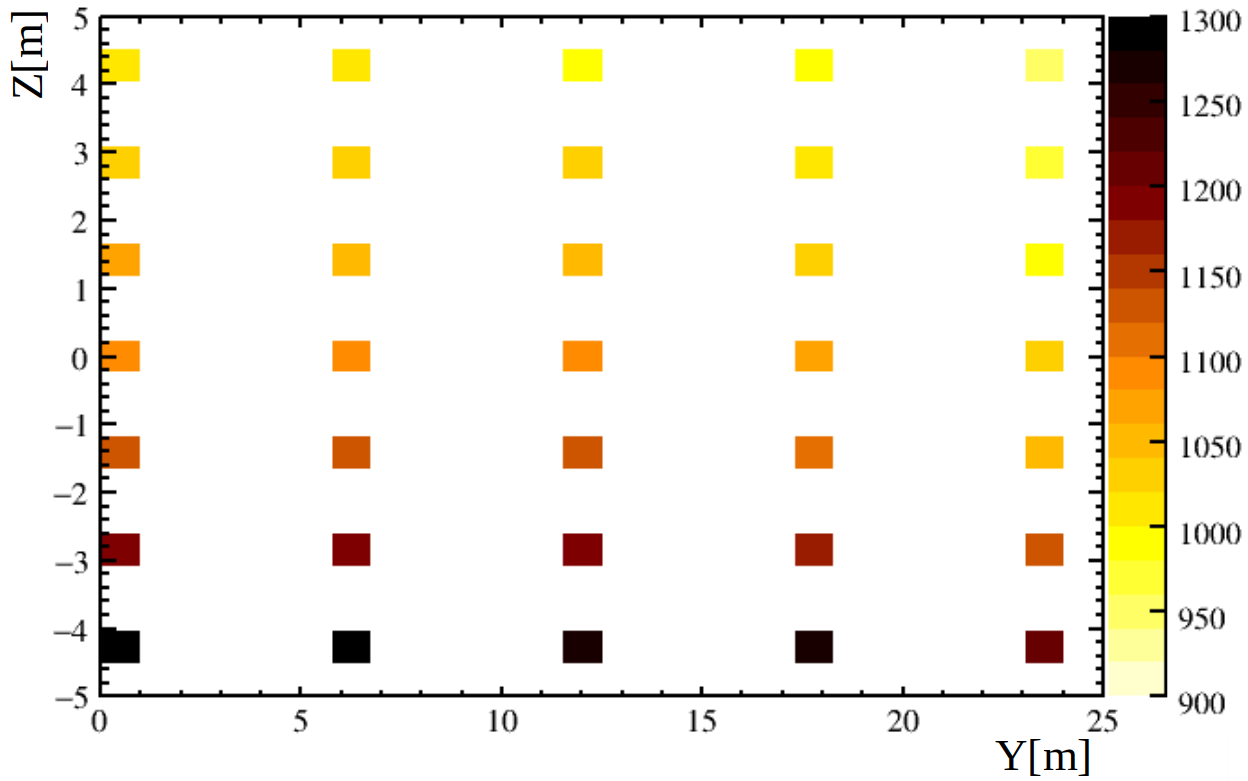}
	\caption{Light yield (total p.e.s per MeV) as a function of true $\upbeta$-particle position in the SLiPS detector - of dimensions shown in figure \ref{PancakeDimensions}. The volume-weighted mean and rms were calculated as 1081 and 43 p.e.s per MeV, respectively.}
	\label{LY_Pancake}
\end{figure}

Energy resolution is an important component in many relevant physics analyses and is ultimately limited by Poisson fluctuations on the total number of photoelectrons recorded by PMTs in a physics event. Efficient light collection is therefore crucial. Figure \ref{LY_Pancake} shows the average detected number of photoelectrons observed in the simulation for 1MeV electrons generated at different z and radial positions in the detector. It can be seen that a high level of light detection are possible with in a SLiPS configuration with a relatively homogeneous response: the variation (rms) in the bulk of the volume is 4$\%$ of the mean and the most extreme light yields lie within 15$\%$. Table \ref{Table_LY} shows comparisons of the typical projected light collection for the SLiPS design with the Borexino, KamLAND, SNO+ and JUNO detectors. The JUNO experiment, also 20kt in mass, will be equipped with $\sim$20,000 large PMTs to achieve their energy resolution goal of 3$\%$ per MeV. This would be a world-leading accomplishment for a kilotonne-scale optical detector. It can be seen that the SLiPS detector could potentially achieve a comparable level of resolution using less than half the number of PMTs and a much simpler construction.

\begin{table}[H]
\caption{Table summarising detector size, PMT coverage and the light collection for the Borexino, KamLAND, SNO+ and JUNO detectors compared to SLiPS.}
\begin{center}
	\begin{tabular}{||c c c c c||}
		\hline
		& Target & PMT & Light coll. & LC/Cov. \\
		& Mass & Coverage & (pe/MeV) & (pe/MeV/\%)\\ \hline \hline
		\makecell{Borexino\\\cite{Borexino}\cite{BorexinoSLIPS}} & 300t & $\sim$30\% & 450 & 15 \\ \hline
		\makecell{KamLAND\\\cite{PhysRevC.85.045504}\cite{KLLightYield}}& 1kT & $\sim$34\% & 200 & 6 \\ \hline
		\makecell{SNO+\\\cite{Lozza:2016rwo}} & 780 & $\sim$50\%& $\sim$520 & 10\\ \hline
		\makecell{JUNO\\\cite{JUNO_DoubleCal}\cite{JUNODesign}}& 20kT & $\sim$80\% & $\sim$1200 & 15 \\ \hline
		SLiPS & 20kT & $\sim$30\% & $\sim$1100 & 37 \\ \hline
\end{tabular}
\end{center}
\label{Table_LY}
\end{table}

\subsection{Position Reconstruction}
High quality position reconstruction is an important part of many physics analyses in order to achieve good signal-background separation. Event positions can be reconstructed using PMT hit times and hit patterns. In the SLiPS detector, the methods of reconstructing vertical and horizontal event positions effectively separate: the compressed and flattened geometry is particularly well suited to reconstruct x and y positions from the PMT hit density, while z-position is better determined from the time separation between direct and reflected wave fronts.

\subsubsection{Reconstruction Using PMT Hit Densities}
\begin{figure}[h]
\centering
\begin{subfigure}{0.49\linewidth}
    \includegraphics[width=\linewidth]{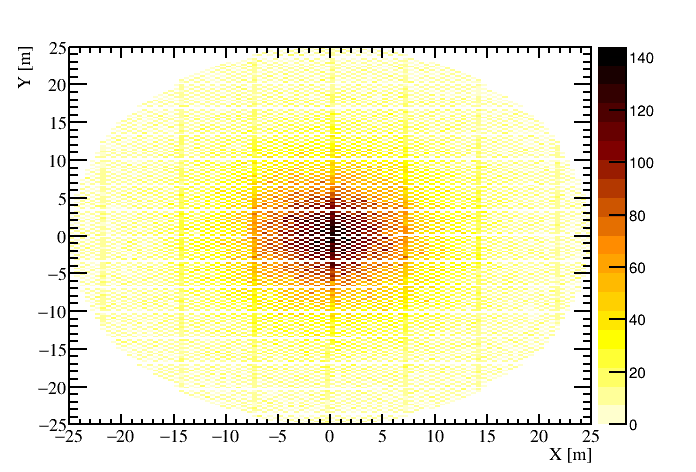}
    \caption{}
    \label{Pancake_Density_0_0}
\end{subfigure}
\begin{subfigure}{0.49\linewidth}
    \includegraphics[width=\linewidth]{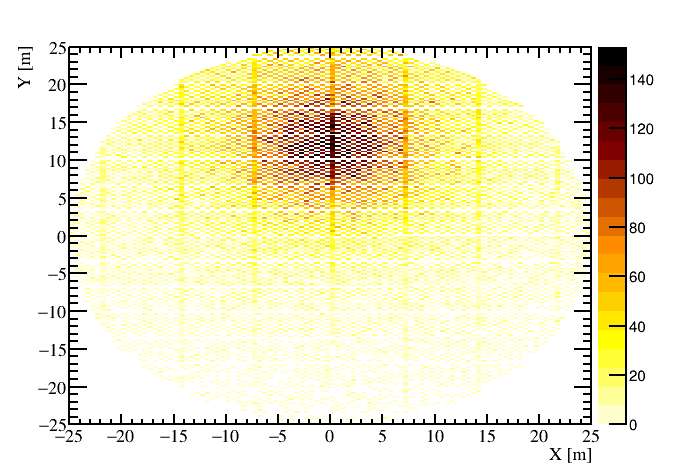}
    \caption{}
    \label{Pancake_Density_0_12}
\end{subfigure}
\begin{subfigure}{0.49\linewidth}
    \includegraphics[width=\linewidth]{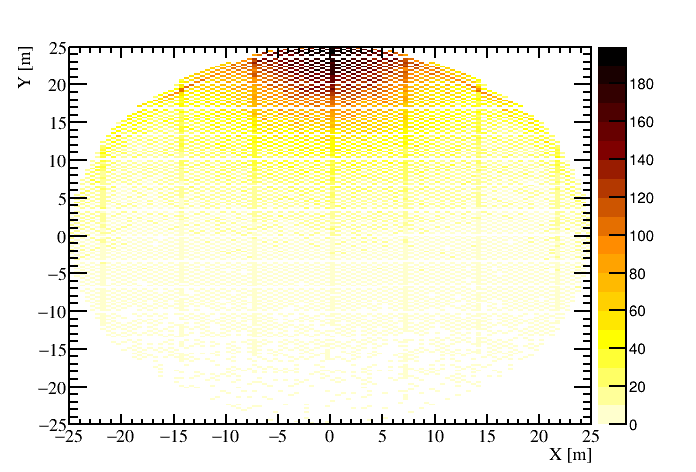}
    \caption{}
    \label{Pancake_Density_0_24p5}
\end{subfigure}
\caption{PMT hit density vs x,y position made with many 3MeV electron events in the Pancake SLiPS detector. Electron events were generated at (a) (0,0,0) (b) (0,12,0)m (c) (0,24.5,0)m. The z-axis reflects arbitrary units.}
\label{Pancake_Density_XY}
\end{figure}

Figure \ref{Pancake_Density_XY} shows PMT hit density plots for multiple 3MeV electron events generated at different y-positions in the 'Pancake' SLiPS detector for x=z=0. These plots serve to illustrate that consideration of the PMT hit distributions alone should allow x and y event positions to be well determined. This capability is also of potential benefit for the use of slow-fluor scintillators used for Cherenkov separation \cite{biller_leming_paton_2020}, where position reconstruction via timing may be compromised to some extent.

To assess this more quantitatively, a likelihood fit was performed to determine event-by-event positions in the x-y plane based on PMT hit densities, using triggered PMT hits occurring within a 400ns time window. Events with z-positions closer to the PMT plane are expected to yield a better x-y resolution owing to a more peaked density profile compared to events closer to the top of the detector. For simplicity, the 'typical' resolution was explored using probability density functions (PDFs) that were generated at various radial displacements for z-positions in the middle of the scintillator. Fits were then performed for 500 3MeV electron events generated at the center of the detector. Results indicate a rms x and y position resolution of between
7cm and 22cm for the 8" PMT configuration, depending on the z-position (with z positions closer to the PMTs having better resolution). This is close to intrinsic expectations based on simple geometric arguments. While information from PMT hit times could also be included to improve the estimation of x-y event positions further, these density-based reconstructions are already approaching a comparable precision to what is typically achieved based on timing in conventional kilotonne-scale LS detectors.


\begin{figure}[h]
	\centering
	\includegraphics[width=68mm]{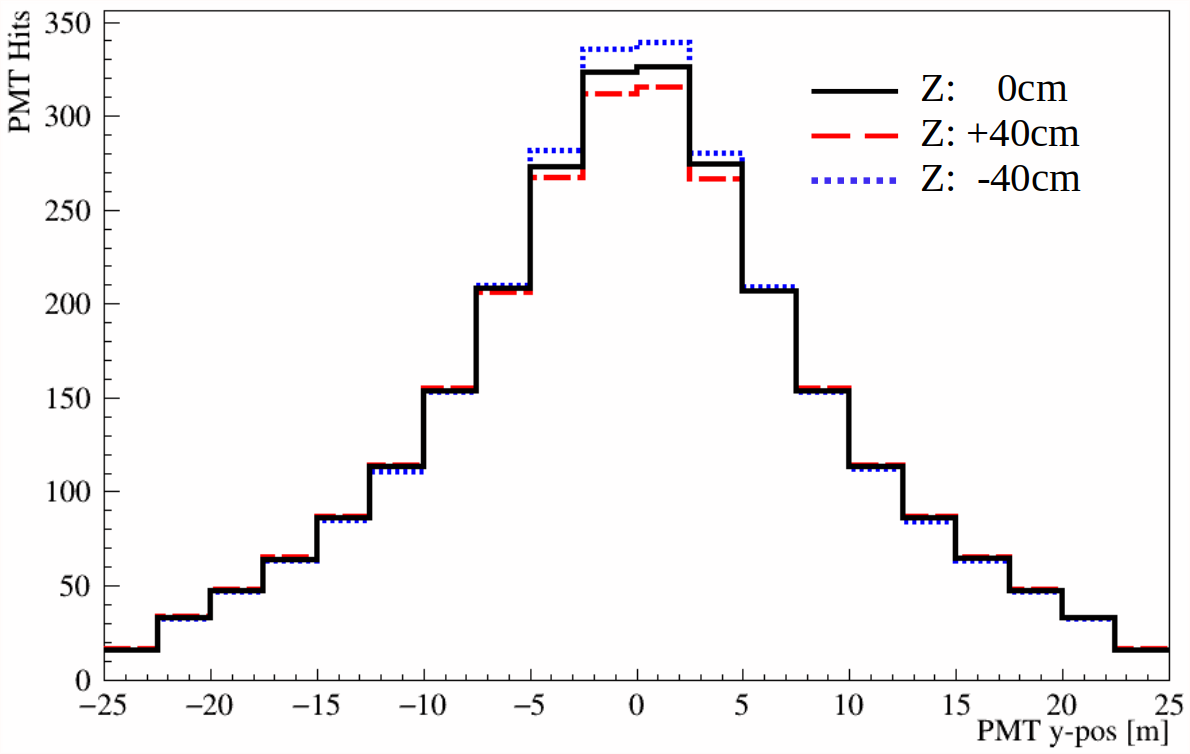}
	\caption{Pdf of PMT hit density vs PMT y-position for 3MeV $\upbeta$ events, for $\upbeta$s generated at z-positions separated by 40cm (x = y = 0). The slight asymmetry in the center of the distribution is due to the binning of hexagonally packed PMT positions in one dimension.}
	\label{HitDensity_zfit}
\end{figure}

As previously indicated, hit densities can also be used to estimate the z-position based on the extent to which the density profile is peaked. While less accurate than x-y localisation, useful constraints can still be obtained. To explore this aspect, a similar procedure as before was followed, but with PDFs of the expected PMT hit density distributions now generated for different z-positions for x=y=0. Figure \ref{HitDensity_zfit} shows representative expected hit density distributions versus PMT y-position (shown in one dimension for clarity). The three PDFs show the expected density distributions for 3MeV electrons at three different z positions separated by 40cm. Differences in the PDF peaks appear subtle, but can be distinguished to a reasonable extent owing to the large photon statistics of each event. The resulting likelihood fits of 500 3MeV events generated at the center of the scintillator yielded a typical standard deviation of 37cm in the z location for 20" PMTs and 30cm for the 8" PMT array. This is more comparable to the vertex resolution of water Cherenkov detectors, but the incorporation of timing information for this axis is expected to lead to a notable improvement in resolution.

\subsubsection{Vertical Position Determination with PMT Hit Times}
\begin{figure}[h]
\begin{subfigure}{0.49\linewidth}
    \includegraphics[width=\linewidth]{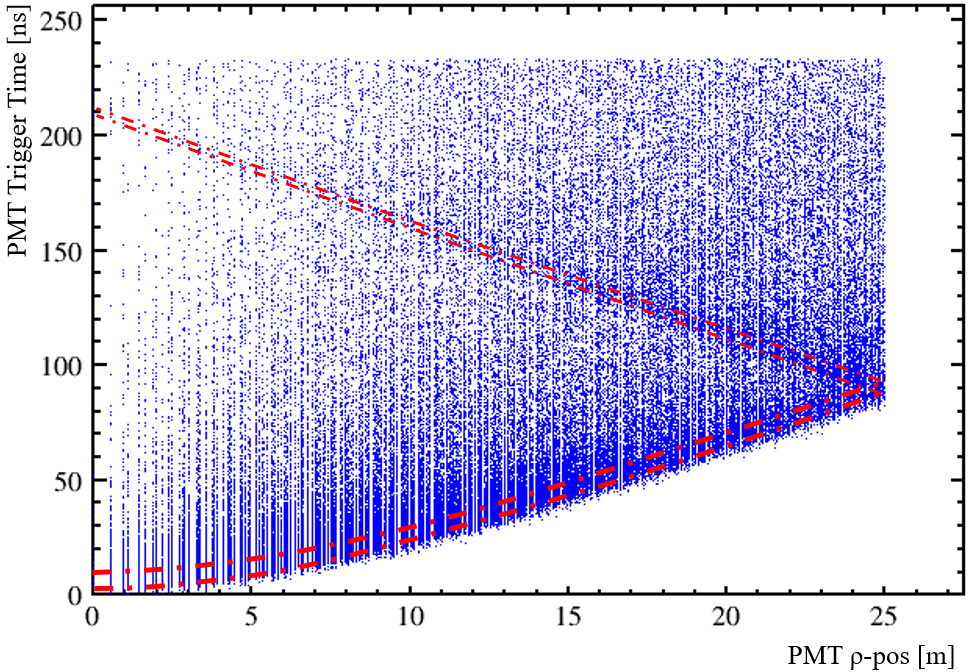}
\caption{}
\end{subfigure}
    \hfill
\begin{subfigure}{0.49\linewidth}
    \includegraphics[width=\linewidth]{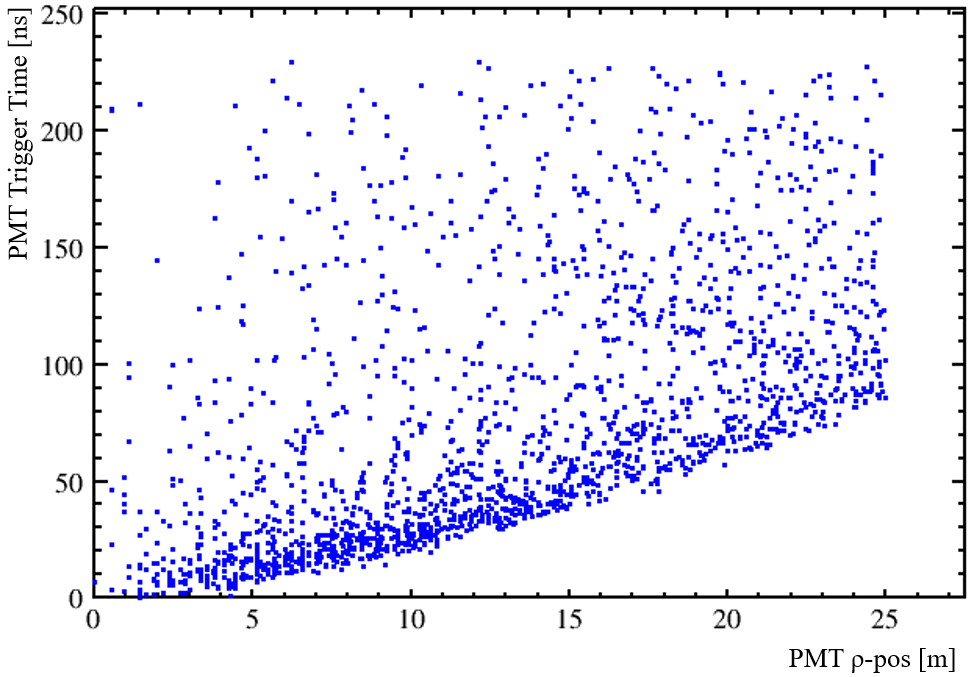}
\caption{}
\end{subfigure}
\begin{subfigure}{0.49\linewidth}
    \includegraphics[width=\linewidth]{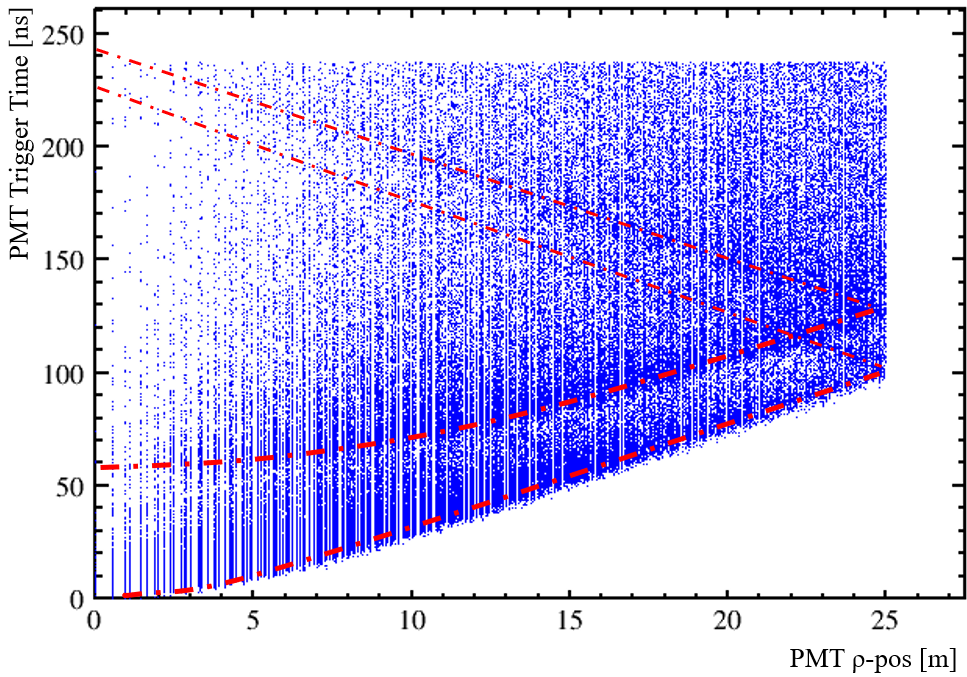}
\caption{}
\end{subfigure}
    \hfill
\begin{subfigure}{0.49\linewidth}
    \includegraphics[width=\linewidth]{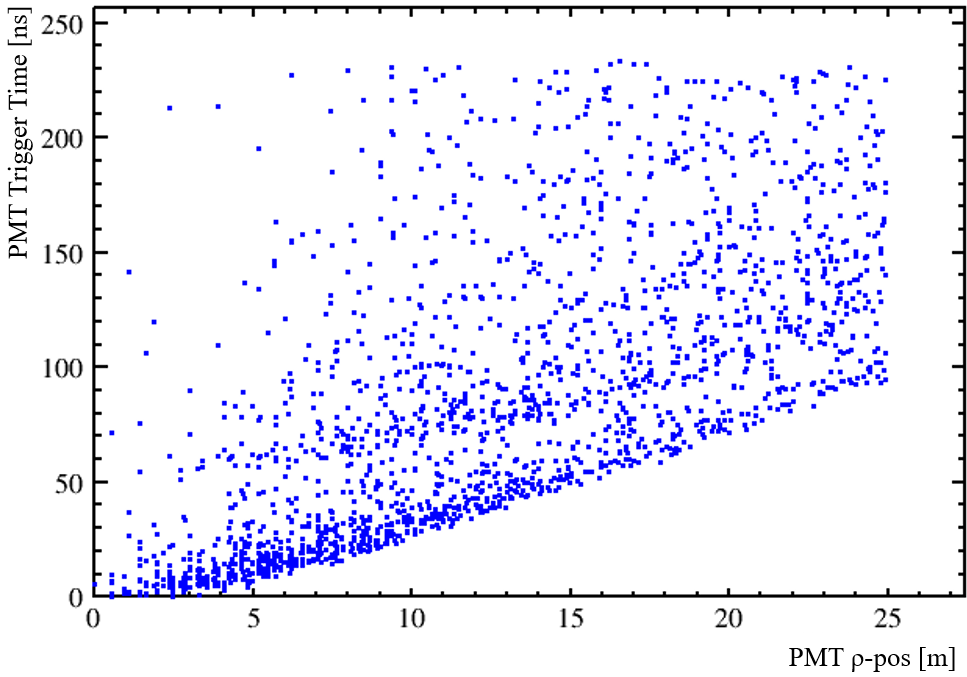}
\caption{}
\end{subfigure}
\begin{subfigure}{0.49\linewidth}
    \includegraphics[width=\linewidth]{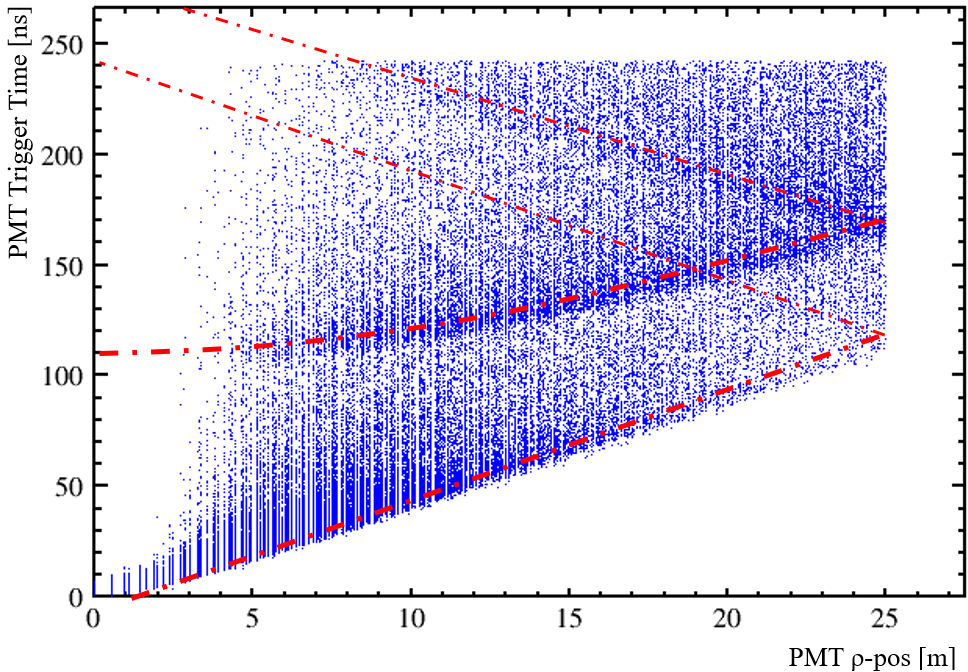}
\caption{}
\end{subfigure}
    \hfill
\begin{subfigure}{0.49\linewidth}
    \includegraphics[width=\linewidth]{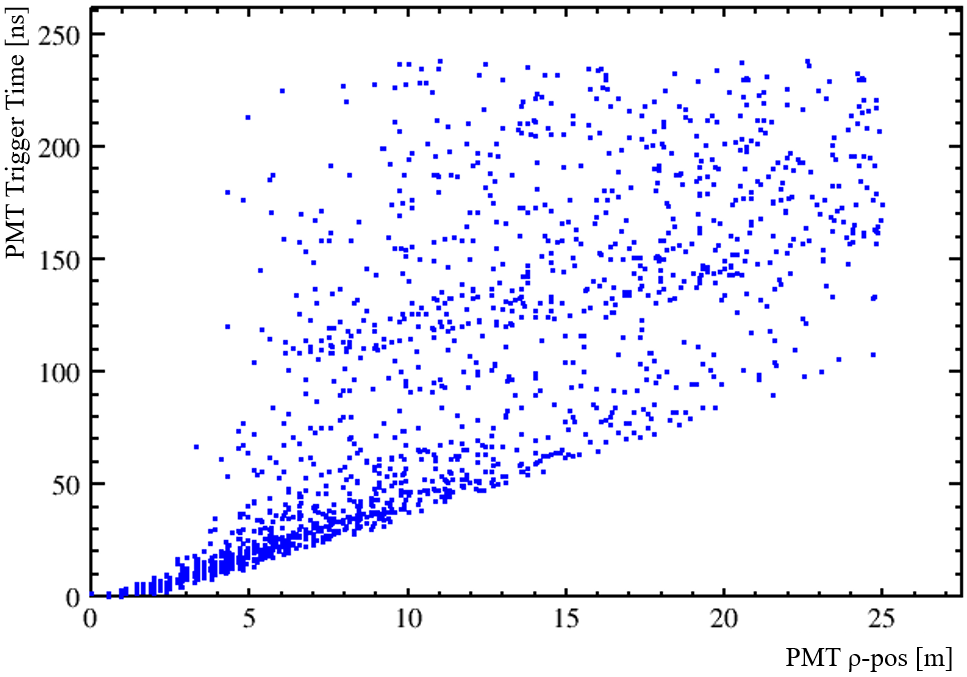}
\caption{}
\end{subfigure}
\caption{PMT hit times vs PMT radial position $\uprho$. Top row: z\textsubscript{ev} = +4.5m, middle row: z\textsubscript{ev} = 0m, bottom row: z\textsubscript{ev} = -4.5m. Left column: Many electron events with analytical calculations of the direct and reflected light wavefronts, right column: an example, single 3MeV electron event.}
\label{HitTimes_v_Pos}
\end{figure}

A strong constraint on the z-position arises from the time differences between direct and reflected scintillation light. Figure \ref{HitTimes_v_Pos} shows triggered PMT hit times vs PMT $\uprho$ (radius in x-y plane) for many electron events generated at an assortment z-positions.

The red dashed lines overlaying the hit times in the first column indicate analytic time-of-flight calculations for direct and reflected light paths of photons traveling in the SLiPS detector. Calculations assumed straight line paths and a single average effective refractive index between that of LAB and ethylene glycol. The calculated light paths include single and double reflections from the ceiling and side walls. The right column of plots demonstrates that the direct and reflected wavefronts are clearly visible on an event-by-event basis. As expected, wavefront separation becomes more difficult for events approaching the reflective sheets. 

Late scintillation emission times produce the hit times seen to lag behind the wavefront regions, where the SNO+ scintillator cocktail of LAB + 2g/L PPO scintillator was assumed to have a leading time decay constant of 4.8ns \cite{SNO+Scintillator}. It will be shown in section \ref{sec_fastpmt_fastscint} how the use of a faster scintillator mixture and/or faster PMTs aids in the reconstruction of event positions. 

\begin{figure}[h]
\centering
\includegraphics[width=0.8\linewidth]{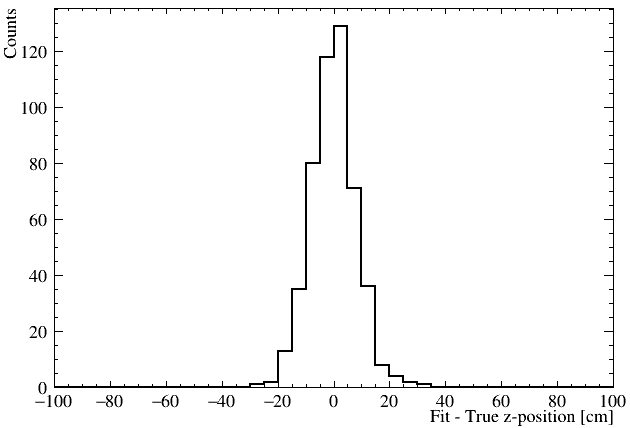}
\caption{Likelihood fit results of the calculated z-position for 500 single 3MeV electron events. The standard deviation of the histogram is 8cm.}
\label{TvRhoDist_Resolution}
\end{figure}

In order to study time-based z-position resolution, a likelihood fit was performed using PDFs based on triggered PMT hit times and positions. In order to compare PMT hit times, a global event trigger time was first estimated. This was approximated by the mean of the 50 earliest PMT hit times, which yielded a standard deviation of $\sim$0.3ns in the trigger times calculated for many events. This global event trigger time was subtracted from all individual PMT trigger times in the following plots. 
PDFs were generated by simulating many 3MeV electron events and binning at z-position intervals separated by 2.5cm. The reconstructed z-position for an event was then taken as that which maximised the likelihood for that event's PMT hit time distribution. 
In order to account for uncertainties in the previously described event trigger time estimation, the trigger time in each fit was allowed to freely float up to $\pm$0.3ns around the estimated event time.

Figure \ref{TvRhoDist_Resolution} shows the fit results for 500 3MeV electron events, all generated at the origin. The standard deviation of the best fit z-positions is found to be 8cm. This is similar to the best position resolution of current LS experiments, despite using fewer PMTs placed in single plane arrangement.

\begin{figure}[h]
\centering
\begin{subfigure}{0.9\linewidth}
    \includegraphics[width=\linewidth]{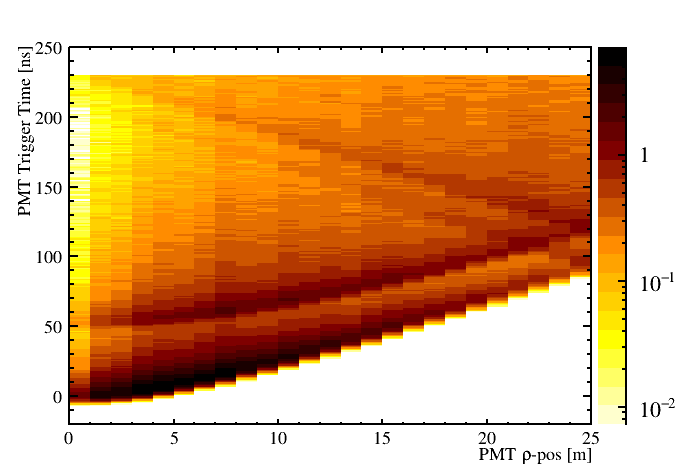}
    \caption{}
    \label{}
\end{subfigure}
    \hfill
\begin{subfigure}{0.9\linewidth}
    \includegraphics[width=\linewidth]{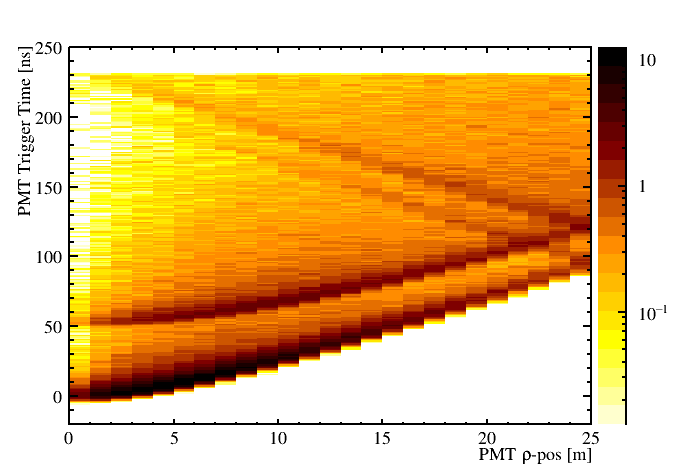}
    \caption{}
    \label{}
\end{subfigure}
\caption{Binned PDFs of PMT hit times vs PMT radial position $\uprho$, for a 3MeV electron event in the center of SLiPS. The plots demonstrate the increased resolution of the direct and reflected wavefronts when both faster PMTs and scintillators are used. (a) 20" r12860 Hamamatsu PMTs (1.3ns TTS) and a 4.8ns leading scintillator emission time constant (b) 8" r5912-100 Hamamatsu PMTs (0.87ns TTS) and a 3ns leading scintillator emission time constant.}
\label{HitTimeRes_v_Pos}
\end{figure}

\subsection{Faster Scintillators \textbf{\&} PMTs}
\label{sec_fastpmt_fastscint}

Figure \ref{HitTimes_v_Pos} shows the wavefront structure that allow for the reconstruction of event position. The resolution of the these wavefront characteristics can be improved by decreasing the spread in PMT hit times. This can be achieved by using faster PMTs and/or scintillator mixtures with faster emission times. 

Figure \ref{HitTimeRes_v_Pos} illustrates the combined impact that PMT TTS and scintillator emission times have on the wavefront resolutions. The pair of plots show wavefront hit time distributions using both faster PMTs as well as faster scintillator emission times, compared to the original SLiPS setup outlined in the introduction. Hamamatsu 8" r5912-100 PMTs were modeled, with the PMT TTS specified in table \ref{Table_PMTs}. These PMTs were placed in a hexagonal close packing configuration as before, with adjacent PMTs spaced by 20.82cm center-to-center (0.5cm glass-to-glass). A total of $\sim$50,000 PMTs were used in the 25m radius Pancake configuration SLiPS detector. The LAB+PPO+bisMSB scintillator cocktail was also altered in simulation such that the leading exponential decay constant for scintillation emission times was changed from 4.8ns to 3ns (achievable through PPO addition \cite{guo_yeh_zhang_cao_qi_wang_chen_2019}).

The side by side comparison shows a noticeable improvement in the wavefront separation. Repeating the time-based likelihood fit on this faster configuration, it was found that z-position resolution was improved from 8cm to 5cm for 3MeV electron events.

\begin{figure}[h]
\centering
\begin{subfigure}{0.44\linewidth}
    \includegraphics[width=\linewidth]{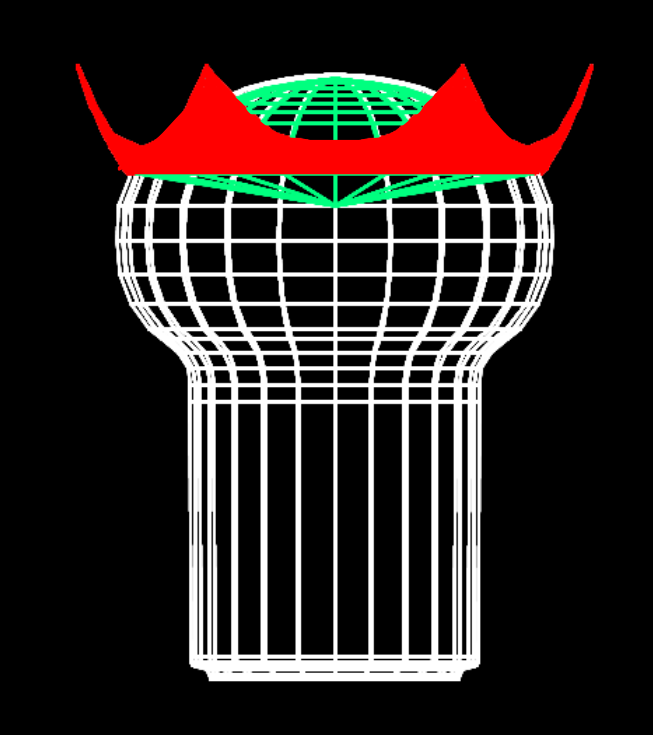}
    \caption{}
    \label{}
\end{subfigure}
    \hfill
\begin{subfigure}{0.49\linewidth}
    \includegraphics[width=\linewidth]{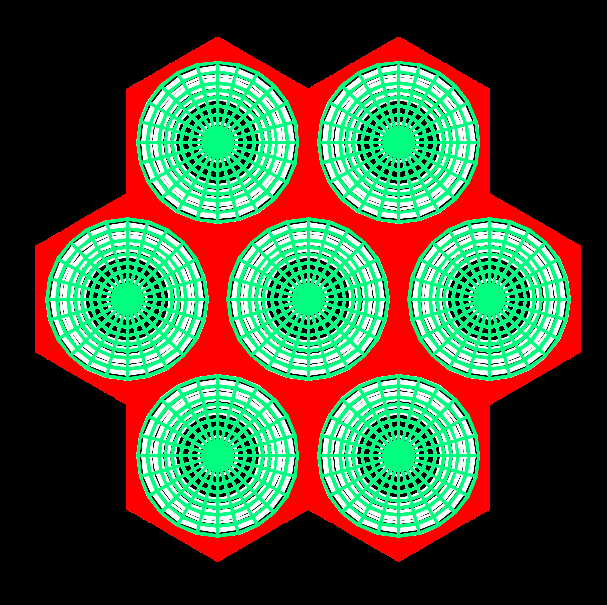}
    \caption{}
    \label{}
\end{subfigure}
\caption{HepRep renderings \cite{HepRep} of the concentrators tested in simulations of the SLiPS detector. (a) Side-on view of a single 8" r5912-100 PMT \cite{kaptanoglu_5912mod_2018}, showing the photocathode (green) and truncated light concentrators (red). (b) Top-down view of a few hexagonally packed PMTs with concentrators masking the regions of the PMT which do not contain a photocathode surface.}
\label{SLIPS_Concentrators}
\end{figure}

\begin{figure}[h]
\centering
\begin{subfigure}{0.9\linewidth}
    \includegraphics[width=\linewidth]{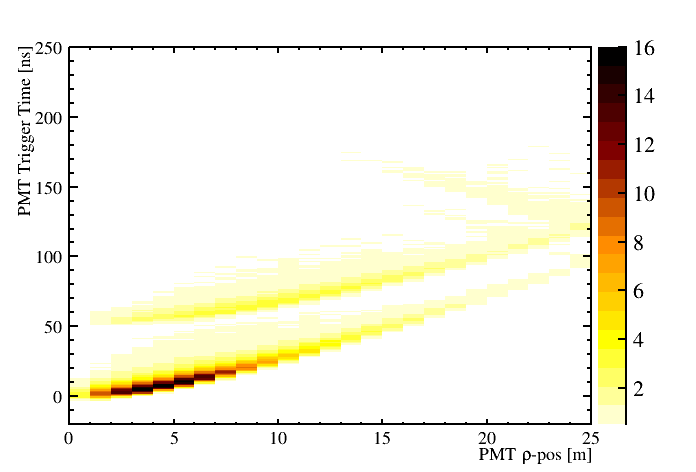}
    \caption{}
    \label{}
\end{subfigure}
    \hfill
\begin{subfigure}{0.9\linewidth}
    \includegraphics[width=\linewidth]{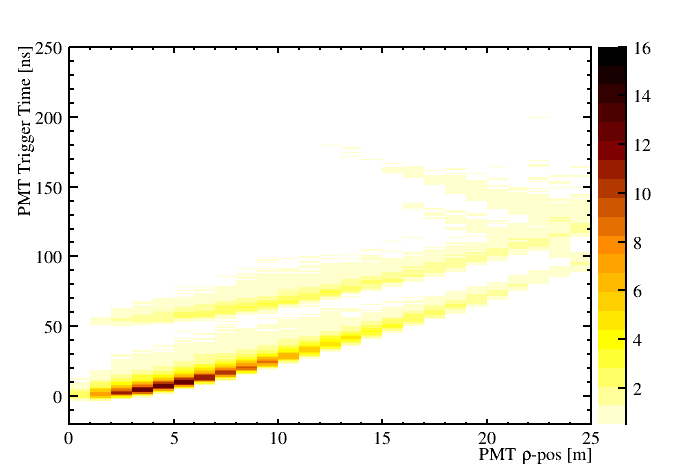}
    \caption{}
    \label{}
\end{subfigure}
\caption{Binned PDFs of PMT hit times vs PMT position for 3MeV electrons at the center of the Pancake SLiPS detector, equipped with fast 8" PMTs and fast 3ns scintillator (a) No light concentrators (b) with light concentrators.}
\label{SLIPS_Concentrators_TvsRho}
\end{figure}

\subsection{Light Concentrators}

In order to increase the light collection and avoid the cost of installing additional PMTs, some scintillator detectors employ light concentrators \cite{Doucas:1996tj}. Generally, the light-detecting photocathode surface does not cover the full width of the PMT itself. Even for very densely packed spherical PMTs, this leaves insensitive regions between PMTs, which reduces the effective photocathode coverage of the detector.

A significant difference between the planar SLiPS detector compared to the common spherical design is that the PMT angular acceptance needs to be as large as possible, since a much higher proportion of the scintillation light in SLiPS arrives at steep angles of incidence. This limits the application of concentrators.
However, modest scale concentrators can still achieve some useful improvement to the photocathode coverage for close-packed hemispherical PMTs by filling in the gaps between active photocathode surfaces.

To investigate this more quantitatively in SLiPS, short light concentrators were modeled in simulation. Figure \ref{SLIPS_Concentrators} shows the shape of the reflectors used to surround each PMT (assuming the 8" r5912-100 PMT). The PMTs were arranged, as before, in a hexagonal close packing arrangement, with a nearest glass-to-glass distance of 0.5cm. To allow for the packed arrangement of PMTs, the concentrators were appropriately truncated. The concentrator base width was made equal to the 9.51cm photocathode radius of the r5912-100 model. Simulations assumed a 90$\%$ specular reflection efficiency for the concentrator surface.

Figure \ref{SLIPS_Concentrators_TvsRho} shows PDFs of PMT hit times vs PMT position for 8" PMTs both with and without concentrators. It can be seen that the concentrators increase the light yield for small $\rho$ (PMTs directly below the event), while the wavefronts become less intense at larger $\rho$. This is predominantly due the steep angles of incidence light arriving at distant PMTs, where there is an increased probability of reflecting back up into the scintillator. For the Pancake SLiPS detector equipped with 8" PMTs, a $\sim$9\% light yield increase from 1180 to 1290 p.e./MeV is observed when concentrators were added. 

\begin{figure}[h]
\includegraphics[width=\linewidth]{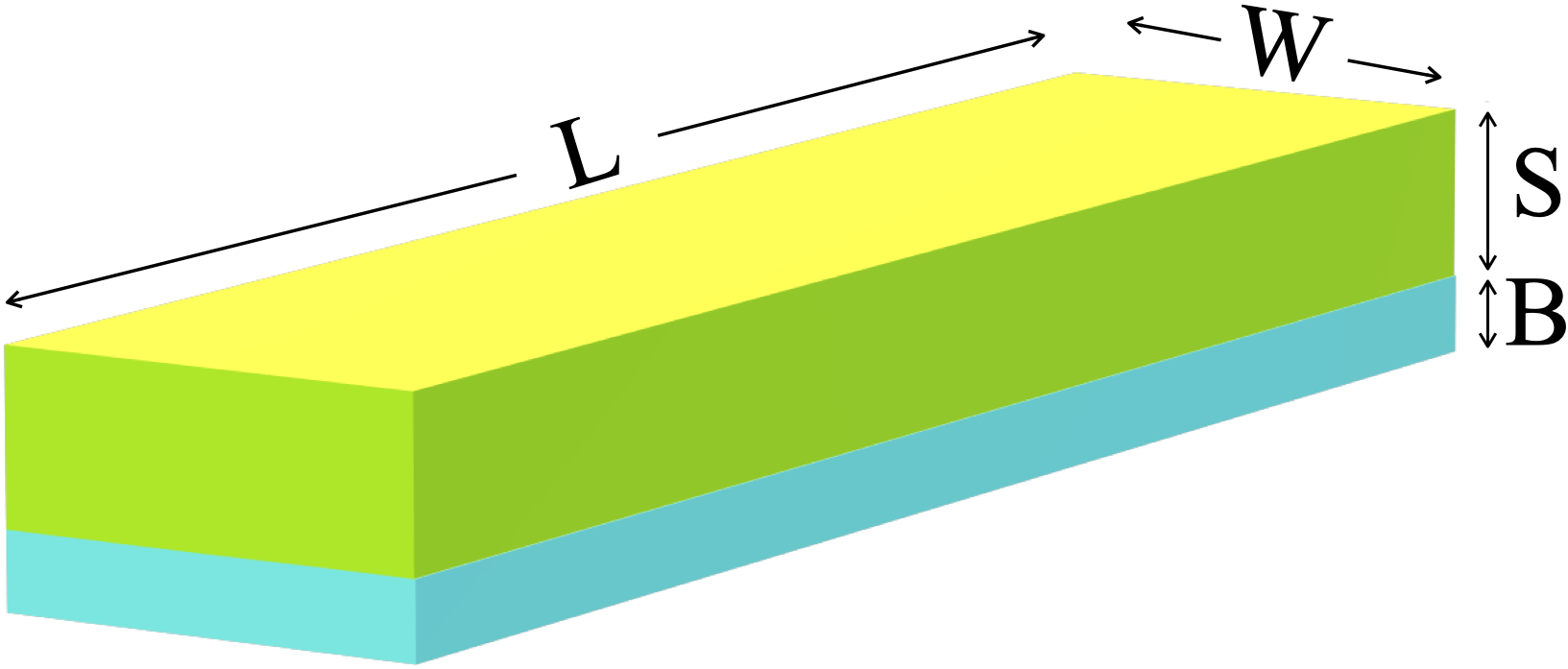}
\caption{Generic box configuration of the SLiPS Detector.}
\label{Cuboid3DDimensions}
\end{figure}

\section{Alternative SLiPS Detector Shape}
\label{sec_cuboid}

The aim of the SLiPS design is to provide a relatively low-cost, large-scale detector of simple construction. It is common for low energy neutrino experiments to be placed in deep underground mines where tunnels are typically excavated for the extraction of geological materials. Therefore, an alternative SLiPS detector shape to consider is a narrow and long box shape. Figure \ref{Cuboid3DDimensions} shows such a generic SLiPS box configuration.

While the box configuration could potentially allow for a simpler and cheaper construction, the narrowness of the detector can complicate position reconstruction due to the increased number of reflections, as well as increasing the radioactive backgrounds from the cavity rock due to the reduced self-shielding. This paper only briefly explores the optical properties of such configurations by looking at the characteristics wavefronts for 3 MeV electrons generated in the middle of such detectors.

\begin{figure}[H]
\centering
\begin{subfigure}{0.8\linewidth}
    \includegraphics[width=\linewidth]{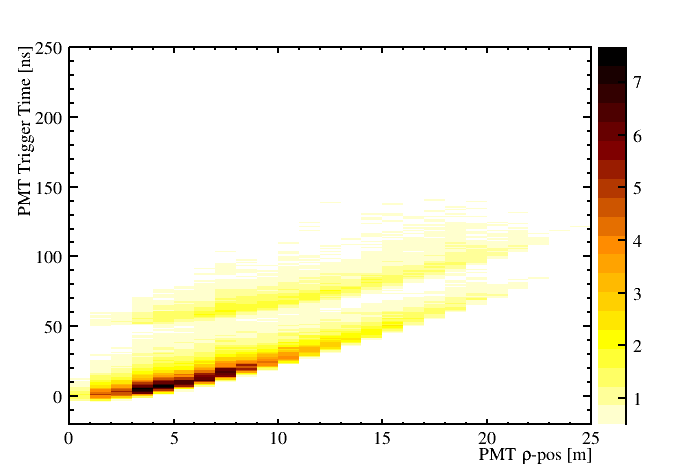}
    \caption{}
    \label{Cuboid_40m_TvsRho}
\end{subfigure} \\
\begin{subfigure}{0.8\linewidth}
    \includegraphics[width=\linewidth]{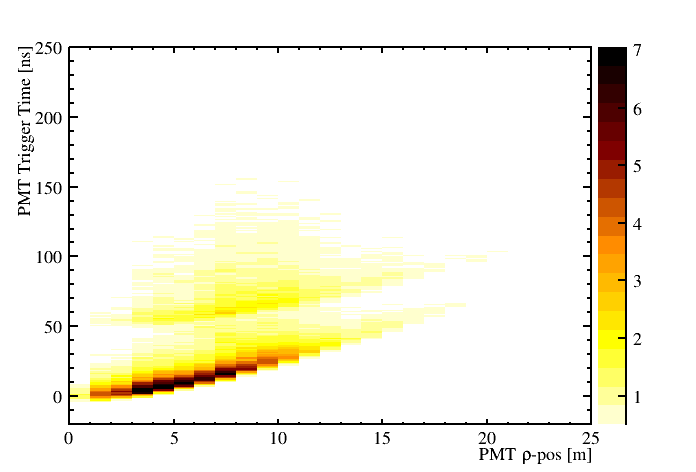}
    \caption{}
    \label{Cuboid_20m_TvsRho}
\end{subfigure} \\
\begin{subfigure}{0.8\linewidth}
    \includegraphics[width=\linewidth]{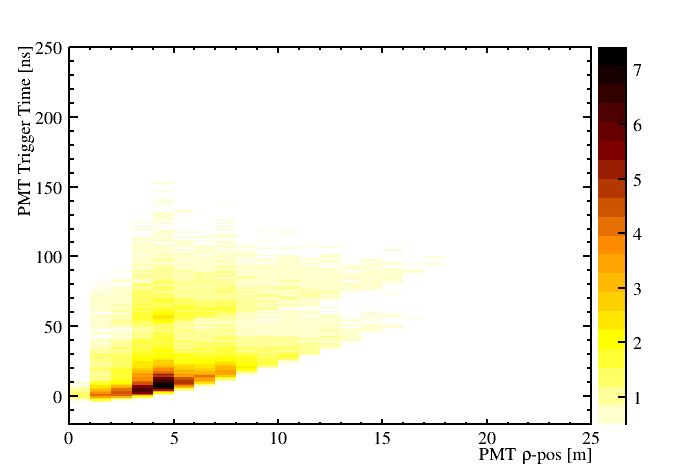}
    \caption{}
    \label{Cuboid_10m_TvsRho}
\end{subfigure}
\caption{Binned PDFs of PMT hit times vs position, generated from many 3MeV electron events at the center of the detector. Plots are for the box configuration of length 100m and width (a) 40m (b) 20m (c) 10m.}
\label{Pancake_vs_Cuboid_TvsRho}
\end{figure}

Figure \ref{Pancake_vs_Cuboid_TvsRho} displays direct and reflected wavefronts for several box configurations. In each case, the length $L$ was chosen to be 100m, the scintillation region height, $S$, was fixed at 10m and the buffer region height, $B$, was fixed at 2m. The width of the box detector was then varied from 10m to 40m for different test cases. For the purposes of demonstration, the wavefronts have again been plotted as PMT hit times vs radial position $\rho$. While more detailed quantitative studies would need to be done for a given configuration, the plots serve to illustrate the potential complications that reflections may have on position reconstruction. As expected, the wavefronts become more spread out and overlap one another as the width of the detector decreases. However, reasonable wavefront separation can still be seen for box widths greater than $\sim$20m (or $W/S > 2$ and $L\gg W$). 

\section{Conclusion}
The SLiPS design offers a highly scalable and simple construction for large scale liquid scintillator detectors, with many potential applications. The use of layered liquids avoids the difficult construction of large transparent barriers to separate scintillation and light detection regions and potentially reduces background contamination from such boundary layers. The studies here also suggest that the use of reflective sheets may yield highly efficient light collection with good position resolution while using far fewer PMTs that those employed in traditional spherical detectors.

This work has been supported by the Science and Technology Facilities Council of the United Kingdom.


\vspace{5.0in}


\renewcommand{\bibname}{Bibliography}
\bibliographystyle{ieeetr}
\bibliography{SLIPS.bib}

\end{document}